\documentclass[reprint,twocolumn]{revtex4}
\usepackage{amsfonts}
\usepackage{amsmath}
\usepackage{amssymb}
\usepackage{charter}
\usepackage{subfigure}
\usepackage{graphicx}
\usepackage{float}

\begin{document}

\title{Multiparameter quantum estimation in a photon system induced by
gravitational redshift}
\author{Wei Ye$^{1}$}
\author{Hui Cao$^{2}$}
\author{Songtao Zhang$^{1}$}
\author{Xiang Zhu$^{1}$}
\author{Huan Zhang$^{3}$}
\thanks{Contact author: zhanghuan@wnu.edu.cn}
\author{Ying Xia$^{3}$}
\author{Shixun You$^{1}$}
\author{Daisheng Zhang$^{1}$}
\author{Shoukang Chang$^{2}$}
\thanks{Contact author: changshoukang@htu.edu.cn}
\affiliation{$^{{\small 1}}$\textit{Jiangxi Provincial Key Laboratory of Low-Altitude Flight Safety and Countermeasures, school of Information Engineering, Nanchang Hangkong University, Nanchang 330063, China}\\
$^{{\small 2}}$\textit{School of Physics, Henan Normal University, Xinxiang
453007, China}\\
$^{{\small 3}}$\textit{Department of Physics, Weinan Normal University,
Wei'nan 714099, China}}

\begin{abstract}
As photons propagate through curved spacetime, gravitational effects become
unavoidable. In particular, gravitational redshift can induce significant
distortion in photon wave packets, making it essential to investigate
parameter estimation within this context. While previous research has
focused on single-parameter estimation using the quantum Cram\'{e}r--Rao
bound, the multiparameter scenario remains largely unexplored. In this work,
we investigate multiparameter quantum estimation for a photon system subject
to gravitational redshift under both amplitude-damping and Ohmic-like
dephasing channels. Our analysis reveals that the quantum Cram\'{e}r--Rao
bound fails to provide a tight error bound for the two-parameter estimation
involving the initial phase and weight parameters in both types of noisy
channels. To overcome this limitation, we numerically compute two tighter
error bounds, i.e., the Holevo Cram\'{e}r--Rao bound and the Nagaoka bound,
when utilizing a semidefinite program. We demonstrate that the Nagaoka bound
yields the tightest error bound among all considered bounds, consistent with
the general hierarchy of multiparameter quantum estimation. Furthermore, for
the three-parameter estimation, including the initial weight parameter, the
phase parameter, and the strength of gravitational redshift, we observe
significantly enhanced estimation precision in the strong-coupling regime
compared to the weak-coupling regime under the amplitude-damping channel.
Similarly, in the Ohmic-like dephasing channel, the sub-Ohmic regime
consistently affords higher precision than the Ohmic and super-Ohmic regimes.
\end{abstract}

\maketitle

\section{Introduction}

Gravitational redshift arising from general relativity is an essential
phenomenon, which provides a direct manifestation of the theoretical
description of spacetime geometry \cite{1,2,3,4} and is also rigorously
substantiated through comprehensive experimental investigations \cite%
{5,6,7,8,9}. Today, the gravitational redshift not only remains a
fascinating physical phenomenon \cite{10,11}, but also serves as a crucial
tool with extensive applications across various branches of physics \cite%
{12,13,14,15}, such as millimetre-scale atomic systems and satellite-based
quantum communications.

When a photon initially prepared by the sender at a specific frequency
traverses curved spacetime, the receiver detects it at a different frequency
due to its exposure to varying local gravitational potentials \cite{16,17,18}%
. Consequently, gravitational redshift affects the realistic photon, which
possesses a finite bandwidth and spatial extension. Recently, a novel method
has been developed to address the question of how the gravitational redshift
affects the quantum state of light. In curved spacetime, two observers
exchanging photons or light pulses are subject to different local
gravitational potentials. The approach involves modeling a photon as a wave
packet of a quantum field propagating through curved spacetime and
establishing a relationship between the wave packet generated by the sender
(typically referred to as Alice) and the one detected by the receiver
(called as Bob) \cite{19,20,21}. The transformation of these wave packets
can be understood as a change in the mode structure of the field, which can
also be interpreted as a change in the basis of the Hilbert space of photons
\cite{22,23,24,25}. Therefore, gravitational redshift can be reinterpreted
either as a unitary rotation of the Hilbert space or, equivalently, as a
multimode mixing operation on the field operator \cite{26,27}.

On the other hand, significant strides have been made in quantum parameter
estimation related to gravitational redshift, encompassing its impact on
phase estimation \cite{16}, precision enhancement via weak measurements \cite%
{17}, and the attenuation of non-Markovian dynamics \cite{18}. Nonetheless,
in these studies, the quantum Cram\'{e}r--Rao bound (CRB) is predominantly
employed to address single-parameter estimation in the context of
gravitational redshift, leaving the more complex area of multiparameter case
largely uncharted. Thus, the exploration of multiparameter estimation in the
context of gravitational redshift remains an open and formidable challenge.
The primary obstacle in multiparameter quantum estimation stems from the
frequent incompatibility of optimal measurements, which often renders the
quantum CRB non-tight \cite{28,29,30,31,32}.

When it comes to the quantum CRB, it can be theoretically formulated by
employing the symmetric logarithmic derivative (SLD) to quantize the
classical CRB, called as the SLD-CRB \cite{28,29,33}, but this quantization
process is not unique \cite{34}. Thus, the right logarithmic derivative
(RLD) provides an alternative approach, similarly called as the RLD-CRB.
Since the corresponding optimal estimators may not align with physically
realizable positive-operator-valued measures (POVMs) \cite{35,36}, the
RLD-CRB is also confronted with potential non-tightness issues.
Consequently, the Holevo Cram\'{e}r--Rao bound (HCRB) that can provide a
tighter precision limit than both the SLD-CRB and RLD-CRB \cite{37} is
widely adopted. In general, the HCRB is achievable through collective
measurements on asymptotically many copies of quantum state \cite{28,29,37},
as well as by single-copy measurements for pure states and displacement
estimation with Gaussian states \cite{38,39,40,41,42}. Despite these
advantages, tthe computational complexity involved in evaluating the HCRB,
which requires solving a sophisticated optimization over a set of
observables, hinders its application in multi-parameter estimation. To
tackle the aforementioned issue, a semidefinite program (SDP) applicable to
finite- and infinite-dimensional Gaussian systems was introduced in Refs.
\cite{42,43}, allowing for straightforward numerical evaluation. Especially,
it is imperative to derive tighter precision bounds that are constrained to
separable, single-copy measurements \cite{44}, because of the prohibitive
experimental burden associated with the collective measurements mandated by
the HCRB. For this purpose, the tight Nagaoka bound (NB) in two-parameter
qubit estimation systems meets this requirement \cite{45}, but the
Nagaoka-Hayashi bound (NHB) fails to generally achieve tightness in
multiparameter cases \cite{46,47,48}. Similarly, the SDP also provides an
effective numerical approach for the complex optimization problems inherent
in computing both NB and NHB \cite{49,50}.

In this paper, we investigate multiparameter quantum estimation for a photon
system subjected to amplitude-damping \cite{51,52,53} and Ohmic-like
dephasing channels \cite{54,55,56,57,58} under the gravitational redshift
effect. For the amplitude-damping channel, we initially examine a
two-parameter estimation scenario involving the initial weight and phase
parameters. Our analysis reveals that the corresponding SLD operators do not
commute, and the mean Uhlmann curvature matrix \cite{28,29,59} is non-zero,
indicating that the SLD-CRB cannot offer a tight error bound, even
asymptotically. Likewise, the RLD-CRB is generally not tight. Consequently,
we numerically compute the HCRB and NB using the SDP. Our results show that
the attainable error bounds (HCRB and NB) achieve higher estimation
precision in the strong-coupling regime compared to the weak-coupling
regime. Remarkably, the NB is the tightest error bound among all bounds,
consistent with the general hierarchy of multiparameter quantum estimation.
Additionally, we explore a three-parameter estimation problem that includes
the weight parameter, phase parameter, and gravitational redshift strength.
Given the complexity of the analytical results, we concentrate on numerical
comparisons of SLD-CRB, RLD-CRB, HCRB, and NHB using the SDP. These
comparisons yield similar observations, i.e., the estimation performance of
the attainable error bounds (HCRB and NHB) is consistently superior under
strong-coupling conditions relative to weak-coupling scenarios. For the
Ohmic-like dephasing channel, we also analyze both two-parameter and
three-parameter estimations. Our results demonstrate that the attainable
error bounds (HCRB, NB, and NHB) consistently exhibit enhanced estimation
precision in the sub-Ohmic regime compared to the Ohmic and super-Ohmic
regimes. Interestingly, for both the amplitude-damping and Ohmic-like
dephasing channels, the RLD-CRB in the three-parameter estimation is
numerically equivalent to the HCRB, suggesting that both RLD-CRB and HCRB
can provide asymptotically tight precision limits.

The structure of this paper is organized as follows. In Sec. II, we provide
a comprehensive review of established findings in multiparameter quantum
estimation theory. In Sec. III, we delve into the spacetime framework
underlying gravitational redshift. In Sec. V, we investigate the
multiparameter estimation challenge by examining a photon system influenced
by gravitational redshift within two distinct quantum channels, involving
the amplitude-damping and Ohmic-like dephasing channels. Finally, our main
conclusions are drawn in the last section.

\section{ Multiparameter quantum estimation}

In this section, we will provide a comprehensive introduction to the
fundamental concepts of multiparameter quantum estimation, establishing the
theoretical foundation essential for our research objectives. Our objective
is to achieve the simultaneous estimation of $d$ unknown parameters $\theta $%
=$(\theta _{1},...,\theta _{d})^{\text{T}}$, which characterize a parametric
family of quantum states $\hat{\rho}_{\theta }.$ In order to estimate these
unknown parameters, we perform the POVM on the quantum state $\hat{\rho}%
_{\theta }.$ The corresponding conditional probability related to
measurement outcome is determined by the Born's rule $P(k|\theta )$=Tr($\hat{%
\rho}_{\theta }\hat{\Pi}_{k}$) where $\hat{\Pi}_{k}$ is the POVM element and
the symbol Tr($\cdot $) is the trace of an operator in Hilbert space. Based
on the obtained measurement outcome, we can invoke a suitable estimator
function $\check{\theta}(k)$ to conjecture the corresponding parameter
values. The performance of the estimator function $\check{\theta}(k)$ is
quantified by the mean square error matrix%
\begin{equation}
\Sigma _{\theta }(\hat{\Pi}_{k},\check{\theta}(k))\text{=}%
\sum\limits_{k}P(k|\theta )(\check{\theta}(k)-\theta )(\check{\theta}%
(k)-\theta )^{\text{T}}.  \label{1}
\end{equation}%
By exerting a locally unbiasedness constraint\textbf{\ }condition%
\begin{eqnarray}
\sum\limits_{k}(\check{\theta}_{\mu }(k)-\theta _{\mu })P(k|\theta ) &=&0,
\notag \\
\sum\limits_{k}\check{\theta}_{\mu }(k)(\left. \partial P(k|\theta )\right/
\partial \theta _{v}) &=&\delta _{\mu v},  \label{2}
\end{eqnarray}%
on the estimator function $\check{\theta}(k),$ one can derive the matrix CRB
for the mean square error matrix \cite{60}
\begin{equation*}
\Sigma _{\theta }(\hat{\Pi}_{k},\check{\theta}(k))\geq F^{-1},
\end{equation*}%
where $F$ is the classical Fisher Information matrix. The matrix CRB serves
as a fundamental theoretical framework that delineates the minimal
achievable mean squared error matrix for a given measurement scheme when
subjected to optimal classical data processing. This theoretical precision
limit can be asymptotically achieved through the implementation of an
appropriate and efficient estimator. In the pursuit of attaining the
ultimate precision limits in multiparameter estimation, the matrix CRB has
been quantized, thereby giving rise to two distinct quantum formulations. A
prominent quantum lower bound for the mean squared error matrix is
intricately linked to the real symmetric quantum Fisher information matrix,
whose elements are given by \cite{61}%
\begin{equation}
J_{uv}^{S}=\frac{1}{2}\text{Tr}\left[ \hat{\rho}_{\theta }(\hat{L}_{u}^{S}%
\hat{L}_{v}^{S}+\hat{L}_{v}^{S}\hat{L}_{u}^{S})\right] ,  \label{3}
\end{equation}%
where the SLD operators $\hat{L}_{u}^{S}$ are determined by the Lyapunov
equation $\partial \hat{\rho}_{\theta }/\partial \theta _{u}$=$\left. (\hat{L%
}_{u}^{S}\hat{\rho}_{\theta }+\hat{\rho}_{\theta }\hat{L}_{u}^{S})\right/ 2$%
. Another crucial lower bound pertains to the RLD quantum Fisher information
matrix, with elements \cite{35,36}
\begin{equation}
J_{uv}^{R}=\text{Tr}\left[ \left( \hat{L}_{u}^{R}\right) ^{\dagger }\hat{\rho%
}_{\theta }\hat{L}_{v}^{R}\right] ,  \label{4}
\end{equation}%
where the RLD operators $\hat{L}_{u}^{R}$ are defined by the relation $%
\partial \hat{\rho}_{\theta }/\partial \theta _{u}$=$\hat{\rho}_{\theta }%
\hat{L}_{u}^{R}.$ The corresponding scalar forms for the SLD-CRB and the
RLD-CRB are respectively given by \cite{28,29}
\begin{eqnarray}
C_{\theta }^{S} &=&\text{tr}[(J^{S})^{-1}],  \notag \\
C_{\theta }^{R} &=&\text{tr}[\text{Re}(J^{R})^{-1}]+\left\Vert \text{Im}%
(J^{R})^{-1}\right\Vert _{1},  \label{5}
\end{eqnarray}%
where the notation tr[$\cdot $] denotes the trace of finite dimensional $%
d\times d$ matrices, $\left\Vert A\right\Vert _{1}$=tr($\sqrt{A^{\dagger }A}$%
) is the trace norm, and Re$\left( \cdot \right) $ and Im$\left( \cdot
\right) $ are the real and imaginary part of a matrix, respectively. In
contrast to the single-parameter estimation, the SLD-CRB is generally not
attainable. This is primarily due to the incompatibility of optimal
measurements for different parameters. Similarly, the RLD-CRB is also
typically not tight, as the optimal estimators associated with the RLD-CRB
may not correspond to physically realizable POVMs. Holevo introduced a
tighter scalar bound known as the HCRB, which is defined as \cite{62}%
\begin{equation}
C_{\theta }^{H}=\min_{\hat{X}}\left[ \text{tr[Re}Z[\hat{X}]\text{]}%
+\left\Vert \text{Im}Z[\hat{X}]\right\Vert _{1}\right] ,  \label{6}
\end{equation}%
where $Z[\hat{X}]$ is a $d\times d$ Hermitian matrix and its matrix elements
are given by $Z[\hat{X}]_{uv}=$Tr$\left[ \hat{\rho}_{\theta }\hat{X}_{u}\hat{%
X}_{v}\right] ,$ in terms of a set of $d$ Hermitian operators $\hat{X}_{u}$
which adhere to the local unbiasedness conditions%
\begin{eqnarray}
\text{Tr}\left[ \hat{\rho}_{\theta }\hat{X}_{u}\right] &=&0,  \notag \\
\text{Tr}\left[ \hat{X}_{u}\partial \hat{\rho}_{\theta }/\partial \theta _{u}%
\right] &=&\delta _{uv}.  \label{7}
\end{eqnarray}%
Typically, the HCRB is tighter than both the SLD-CRB and RLD-CRB. More
recently, it has been proven that the discrepancy between the HCRB and the
SLD-CRB is bounded by a factor two \cite{28,29}
\begin{equation}
C_{\theta }^{S}\leq C_{\theta }^{H}\leq C_{\theta }^{U}\leq 2C_{\theta }^{S},
\label{8}
\end{equation}%
where the first upper bound can be expressed as $C_{\theta }^{U}$=$C_{\theta
}^{S}+\left\Vert (J^{S})^{-1}D(J^{S})^{-1}\right\Vert _{1}$ and $D$ is the
mean Uhlmann curvature matrix \cite{50}, whose elements%
\begin{equation}
D_{uv}=-\frac{i}{2}\text{Tr[}\hat{\rho}_{\theta }[\hat{L}_{u}^{S},\hat{L}%
_{v}^{S}]\text{].}  \label{9}
\end{equation}%
It is worth noting that the SLD-CRB is fully tight in the scenario where the
SLD operators $\hat{L}_{u}^{S}$ and $\hat{L}_{v}^{S}$ commute, i.e., $[\hat{L%
}_{u}^{S},\hat{L}_{v}^{S}]$=$0,$ for all $u,v,$ and the equality $C_{\theta
}^{S}$=$C_{\theta }^{H}$ holds. In such cases, a set of common eigenstates
associated with these commuting SLD operators can always be identified.
These eigenstates can be employed as the POVM basis, enabling the saturation
of the SLD-CRB through single-copy measurements. Furthermore, the SLD-CRB is
asymptotically tight, i.e., there are some special quantum states that
satisfy $[\hat{L}_{u}^{S},\hat{L}_{v}^{S}]\neq 0$ and the mean Uhlmann
curvature matrix $D$ is a zero matrix. Additionally, it can be demonstrated
that the equality $C_{\theta }^{S}$=$C_{\theta }^{H}$ can be asymptotically
achieved through the application of collective measurements.

Despite its theoretical promise, the practical implementation of collective
measurements continues to pose significant experimental challenges within
the constraints of current technological capabilities. Recognizing this
limitation, Nagaoka pioneered the development of a more informative scalar
bound specifically tailored for two-parameter estimation scenarios \cite{45}
\begin{eqnarray}
C_{\theta }^{N} &=&\min_{\hat{X}}\{\text{Tr}[\hat{\rho}_{\theta }\hat{X}_{1}%
\hat{X}_{1}+\hat{\rho}_{\theta }\hat{X}_{2}\hat{X}_{2}]  \notag \\
&&+\text{TrAbs}[\hat{\rho}_{\theta }[\hat{X}_{1},\hat{X}_{2}]]\},  \label{10}
\end{eqnarray}%
where TrAbs$[\hat{K}]$ represents the sum of the absolute values of the
eigenvalues of the operator $\hat{K}.$ The NB has been demonstrated to
constitute a tight scalar bound specifically for two-parameter estimation
scenarios. To extend its applicability to higher-dimensional parameter
spaces, Conlon \textit{et al.} developed a comprehensive generalization of
the NB, termed the NHB, which is expressed as \cite{44}

\begin{eqnarray}
C_{\theta }^{N} &=&\min_{\hat{L},\text{ }\hat{X}}\{\left. \mathbf{Tr[}\hat{S}%
_{\theta }\hat{L}\mathbf{]}\right\vert \hat{L}_{uv}\text{=}\hat{L}_{vu}\text{
Hermitian,}  \notag \\
&&\hat{L}\geq \hat{X}\hat{X}^{\text{T}}\},  \label{11}
\end{eqnarray}%
where $\hat{S}_{\theta }$=$1_{d}\otimes \hat{\rho}_{\theta }$ exists in an
expanded classical-quantum Hilbert space$,$ $1_{d}$ is\textbf{\ }the $%
d\times d$ identity matrix, $\hat{L}$ is the $d\times d$ matrix of Hermitian
operators, the symbol $\mathbf{Tr}$[$\cdot $] denotes\textbf{\ }the trace
over both classical and quantum systems.

To streamline our mathematical presentation, we adopt the unified notation $%
C_{\theta }^{N}$ to represent both the NB and its generalized counterpart,
the NHB. It is worth noting that Gill and Massar independently developed an
alternative bound; however, given its typically weaker performance in terms
of tightness when compared to the NHB, we have deliberately omitted it from
the scope of our current analysis \cite{63}. From a theoretical perspective,
the most informative bound can be rigorously characterized as the minimal
scalar CRB obtained through optimization across the complete space of all
possible POVMs \cite{44,64,65}%
\begin{equation}
C_{\theta }^{MI}=\min_{\text{POVM}}[\text{tr}[F^{-1}]],  \label{12}
\end{equation}%
which satisfies the following chain of inequalities
\begin{eqnarray}
\text{tr}[\Sigma _{\theta }(\hat{\Pi}_{k},\check{\theta}(k))] &\geq
&C_{\theta }^{MI}\geq C_{\theta }^{N}\geq C_{\theta }^{H}  \notag \\
&\geq &\max [C_{\theta }^{S},C_{\theta }^{R}].  \label{13}
\end{eqnarray}%
It is crucial to highlight that in the single-parameter estimation, the
SLD-CRB, HCRB, NHB, and the most informative bound yield identical numerical
results. In the context of two-parameter estimation, the NB serves as a
tight scalar bound, demonstrating equivalent numerical outcomes to the most
informative bound. Additionally, when estimating any number of parameters
using pure quantum states, the HCRB and NHB exhibit numerical equality \cite%
{44,65}.

\begin{figure}[tbp]
\label{Fig1} \centering\includegraphics[width=\columnwidth]{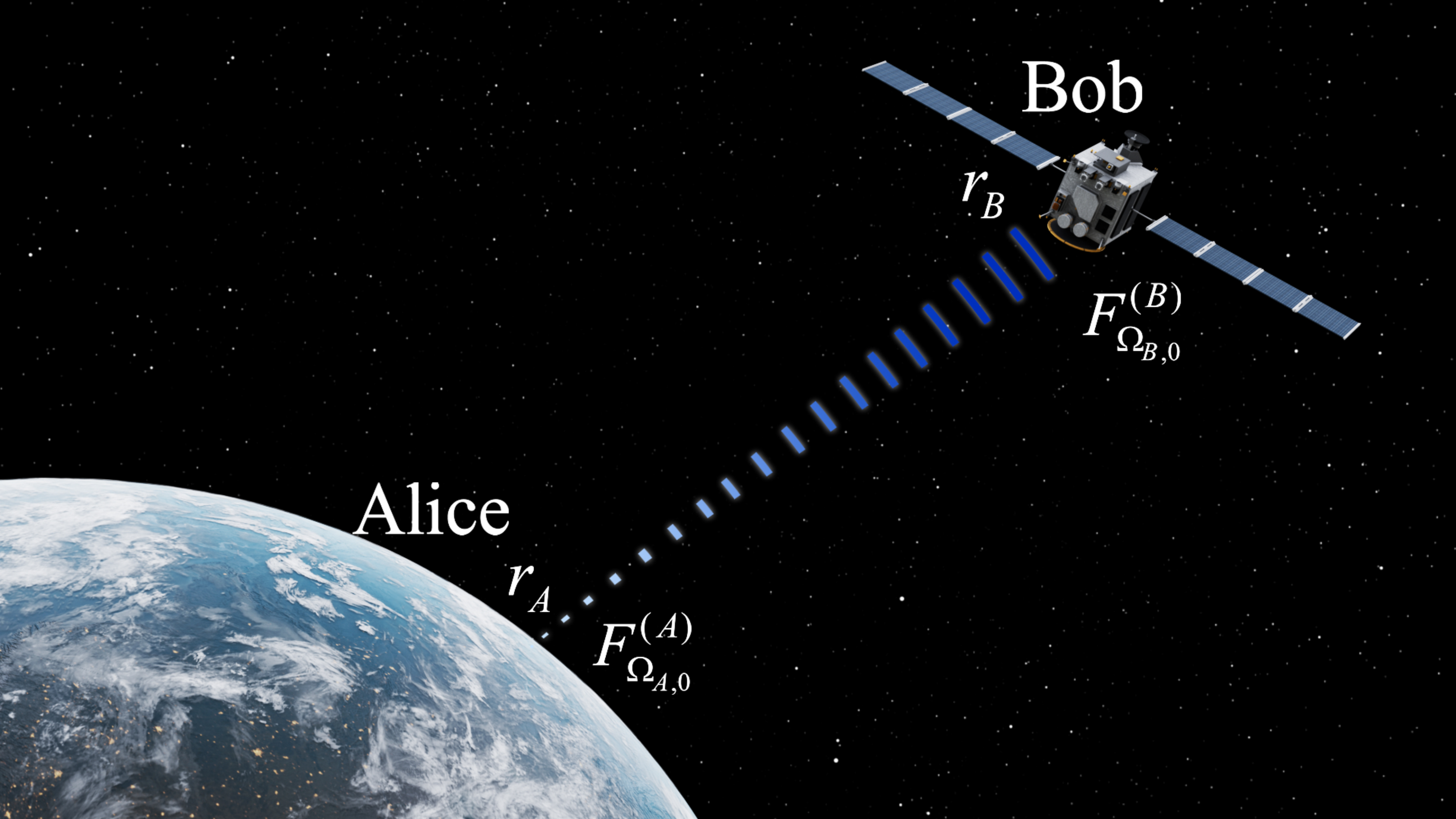}\newline
\caption{{}(Color online) The sender, Alice, located at position $r_{A}$,
transmits a photonic wave packet initially characterized by a frequency
distribution $F_{\Omega _{A,0}}^{(A)}$. This wave packet propagates through
a curved spacetime background and is subsequently received by Bob at
position $r_{B}$, with the frequency distribution $F_{\Omega _{B,0}}^{(B)}$.
}
\end{figure}

\section{Spacetime background of gravitational redshift}

In this section, we begin by introducing the spacetime background under
consideration. For analytical simplicity, we adopt a non-rotating planetary
model and describe the external spacetime geometry using the $(3+1)$%
-dimensional Schwarzschild spacetime, which exhibits both spherical symmetry
and static characteristics \cite{2}. The corresponding Schwarzschild vacuum
metric $g_{\mu v}$ can be expressed as%
\begin{equation}
g_{\mu v}=\text{diag}\left( -f(r),\frac{1}{f(r)},r^{2},r^{2}\sin \theta
\right) ,  \label{14}
\end{equation}%
where $f(r)=1-r_{s}/r,$ and $r_{s}=2M$ represents the Schwarzschild radius
of the planet with $M$ being the mass of the planet. We mainly focus on the
transmission of the wavepacket from Earth to a receiver located at a
specific distance. The corresponding gravitational effect primarily depends
on the Schwarzschild radius $r_{s}.$ Typically, a photon can be described as
a wavepacket with a frequency distribution $F_{\Omega _{K,0}}^{(K)}$ \cite%
{13,14}$.$ The annihilation operator of the photon, as described by
observers at different locations, is given by%
\begin{equation}
\hat{a}_{\Omega _{K,0}}(\tau _{K})=\int_{0}^{+\infty }d\Omega
_{K}e^{-i\Omega _{K}\tau _{K}}F_{\Omega _{K,0}}^{(K)}(\Omega _{K})\hat{a}%
_{\Omega _{K}},  \label{15}
\end{equation}%
where the subscripts $K=A$ and $K=B$ correspond to the observers Alice and
Bob, respectively, $\Omega _{K}$ denotes the frequency of the photon as
measured locally by the observer $K$ at proper time $\tau _{K},$ and $\Omega
_{K,0}$ is the peak frequency of the frequency distribution $F_{\Omega
_{K,0}}^{(K)}.$ The operator $\hat{a}_{\Omega _{K,0}}$ adheres to the
canonical commutation relation $[\hat{a}_{\Omega _{K,0}},\hat{a}_{\Omega
_{K,0}}^{\dagger }]=\delta (\Omega _{K}-\Omega _{K}^{\prime }).$ At proper
time $\tau _{A}$ and location $r_{A},$ the sender Alice transmits a wave
packet $F_{\Omega _{A,0}}^{(A)}$ to the receiver Bob. The wave packet
propagates through curved spacetime and is detected as $F_{\Omega
_{B,0}}^{(B)}$ by Bob at proper time $\tau _{B}=\Delta \tau +\sqrt{\left.
f(r_{B})\right/ f(r_{A})}\tau _{A}$ and location $r=r_{B}>r_{A}.$ Based on
the relation between the annihilation operators $\hat{a}_{\Omega _{A,0}}$
and $\hat{a}_{\Omega _{B,0}}$ \cite{13,14}$,$ the connection between
frequency distributions $F_{\Omega _{K,0}}^{(K)}$ in different reference
frames is established as follows%
\begin{equation}
F_{\Omega _{B,0}}^{(B)}(\Omega _{B})=\sqrt[4]{\frac{f(r_{B})}{f(r_{A})}}%
F_{\Omega _{A,0}}^{(A)}\left( \sqrt{\frac{f(r_{B})}{f(r_{A})}}\Omega
_{B}\right) .  \label{16}
\end{equation}%
Because of spacetime curvature and the effects of gravitational potential,
the wave packet transforms as it travels. In the scenarios, we examine, $%
r_{B}>r_{A},$ the frequency $\Omega _{B}$ observed by Bob is lower than the
frequency $\Omega _{A}$ emitted by Alice, a phenomenon known as redshift.
When Alice generates a sharp frequency mode $\Omega _{A},$ Bob receives it
at the shifted frequency $\Omega _{B}$, whose pictorial representation can
be seen in Fig. 1. The connection between $\Omega _{B}$ and $\Omega _{A}$
can be derived by solving the eigenvalue equation for the relevant modes,
i.e.,%
\begin{equation}
\chi ^{2}=\frac{\Omega _{B}}{\Omega _{A}}=\sqrt{\frac{f(r_{B})}{f(r_{A})}},
\label{17}
\end{equation}%
which is the widely recognized formula for gravitational redshift and
satisfies $\chi >1$ \cite{66}$.$ As described in Refs. \cite{16,17,18}, the
propagation of a photon between two distinct locations in curved spacetime
can be likened to a beam-splitter operation acting on the photon as it
travels. Significantly, the operator in Eq. (\ref{15}) is capable of
describing the same optical mode at two different points before and after
the propagation process. Therefore, the mode $\hat{a}_{\omega _{0}}^{\prime
} $ could be decomposed into the mode $\hat{a}_{\omega _{0}}$ and orthogonal
mode $\hat{a}_{\perp },$ which is described as%
\begin{equation}
\left(
\begin{array}{c}
\hat{a}_{\omega _{0}}^{\prime } \\
\hat{a}_{\perp }^{\prime }%
\end{array}%
\right) =\left(
\begin{array}{cc}
\cos \theta & e^{i\varphi }\sin \theta \\
-e^{-i\varphi }\sin \theta & \cos \theta%
\end{array}%
\right) \left(
\begin{array}{c}
\hat{a}_{\omega _{0}} \\
\hat{a}_{\perp }%
\end{array}%
\right) ,  \label{18}
\end{equation}%
where the angle $\theta $ and phase $\varphi $ are determined by the overlap
of two modes, i.e., $\cos \theta (\chi )=\left\vert \left\langle 1_{\omega
_{0}}^{\prime }|1_{\omega _{0}}\right\rangle \right\vert $ and $\varphi
(\chi )=\arg (\left\langle 1_{\omega _{0}}^{\prime }|1_{\omega
_{0}}\right\rangle ),$ with $\left\vert 1_{\omega _{0}}\right\rangle =\hat{a}%
_{\omega _{0}}^{\dagger }\left\vert 0\right\rangle $ and $\left\vert
1_{\omega _{0}}^{\prime }\right\rangle =\hat{a}_{\omega _{0}}^{\prime
\dagger }\left\vert 0\right\rangle .$ For the convenience of the following
discussion and analysis, we set the phase $\varphi (\chi )=0$ \cite{25}$.$
Due to the curvature of spacetime and the influence of gravitational
potential, the wave packet undergoes deformation during its propagation. The
channel between Alice and Bob (i.e., spacetime itself) exhibits noise
characteristics, and its transmission quality can be quantified by fidelity $%
\tilde{F}=\left\vert \Theta \right\vert ^{2}$%
\begin{equation}
\Theta =\int_{0}^{+\infty }d\Omega _{B}F_{\Omega _{B,0}}^{(B)\ast }(\Omega
_{B})F_{\Omega _{A,0}}^{(A)}(\Omega _{B}),  \label{19}
\end{equation}%
where $\Theta $ denotes the wave packet overlap between the distributions $%
F_{\Omega _{B,0}}^{(B)}(\Omega _{B})$ and $F_{\Omega _{A,0}}^{(A)}(\Omega
_{B})$. For a perfect channel, the condition $\left\vert \Theta \right\vert
=1$ holds. This can be expressed equivalently through the relations: $\cos
\theta =\Theta $ and $\sin ^{2}\theta =1-\Theta ^{2}$ where $\theta $ lies
in the interval $\left[ 0,\pi /2\right) .$ When selecting a wave packet
characterized by a Gaussian distribution of the form $F_{\Omega _{0}}(\Omega
)=\left. e^{-(\Omega -\Omega _{0})^{2}/4\sigma ^{2}}\right/ \sqrt[4]{2\pi
\sigma ^{2}}$ with $\sigma $ being the wave packet width \cite{13,14}, one
can derive the following results%
\begin{equation}
\cos \theta =\sqrt{\left. 2\chi ^{2}\right/ (1+\chi ^{4})}e^{-\frac{(\chi
-1)^{2}\Omega _{B,0}^{2}}{4(1+\chi ^{4})\sigma ^{2}}.}  \label{20}
\end{equation}%
In typical communication scenarios with $\Omega _{B,0}=700$ THz and $\sigma
=1$ MHz, when Bob is located at a significant distance from Earth ($\chi
-1=3.5\times 10^{-10}$), the influence of gravity on photon propagation
becomes non-negligible ($\left. (\chi -1)^{2}\Omega _{B,0}^{2}\right/
4(1+\chi ^{4})\sigma ^{2}\thicksim 7.5\times 10^{-3}$) \cite{16,17,18}. In
the scenario where Alice and Bob are in flat spacetime or are situated at
the same altitude, the communication channel between them operates
flawlessly without any influence from gravitational redshift $(\chi =1)$.
Under these conditions, the fidelity reaches its maximum value, with $\cos
^{2}\theta =1$. However, as the gravitational redshift effect becomes more
significant ($r_{B}>r_{A}$), the fidelity of the channel decreases, leading
to $\cos \theta <1$. Here, $\theta $ serves as a measure of the
gravitational redshift: $\theta $ $=0$ indicates the absence of redshift,
corresponding to perfect signal overlap, while $\theta $ = $\pi /2$
represents the maximum redshift effect, resulting in complete signal
mismatch. In the subsequent analysis, we utilize $\theta $ to quantify the
strength of gravitational redshift, with larger values of $\theta $
indicating a more substantial redshift effect \cite{16,17,18}.

\section{Multiparameter quantum estimation in two distinct quantum channels}

In this section, we evaluate the ultimate bounds for multiparameter quantum
estimation through analyzing a photon system subjected to gravitational
redshift effects in two distinct quantum channels, include the
amplitude-damping channel and the Ohmic-like dephasing channel. Through this
analysis, we establish the corresponding five fundamental precision limits:
the SLD-CRB, RLD-CRB, HCRB, NB, and NHB.

\subsection{Amplitude-damping channel}

Here we focus on a photon system coupled with a structured reservoir at zero
temperature, where the reservoir is initially prepared in a vacuum state.
The complete dynamics of this coupled system is determined by the total
Hamiltonian \cite{51,52,53}
\begin{equation}
\hat{H}=\omega _{0}\hat{a}^{\dagger }\hat{a}+\sum\limits_{k}\omega _{k}\hat{b%
}_{k}^{\dagger }\hat{b}_{k}+\sum\limits_{k}g_{k}(\hat{b}_{k}\hat{a}^{\dagger
}+\hat{b}_{k}^{\dagger }\hat{a}),  \label{21}
\end{equation}%
where $\omega _{0}$ represents the transition frequency of the photon
system, $\hat{a}^{\dagger }$ ($\hat{a})$ denotes the creation (annihilation)
operator for the photon mode, $\hat{b}_{k}^{\dagger }$ ($\hat{b}_{k})$
signifies the creation (annihilation) operator for the field mode $k$ with
frequency $\omega _{k},$ and $g_{k}$ is the coupling constant between the
photon and the reservoir associated with mode $k.$ The nonunitary generator
of the reduced photon system can be given by \cite{51,52,53}%
\begin{equation}
\text{\L }_{A}(\hat{\rho}_{t})=\frac{\gamma _{t}}{2}(2\hat{a}\hat{\rho}_{t}%
\hat{a}^{\dagger }-\hat{a}^{\dagger }\hat{a}\hat{\rho}_{t}-\hat{\rho}_{t}%
\hat{a}^{\dagger }\hat{a}),  \label{22}
\end{equation}%
where $\gamma _{t}$ denotes the time-dependent decay rate. With a single
excitation present in the entire system, the structure of the reservoir can
be effectively characterized by a Lorentzian spectral density
\begin{equation}
J(\omega )=\frac{1}{2\pi }\frac{\gamma _{0}\lambda }{(\omega _{0}-\omega
)^{2}+\lambda ^{2}},  \label{23}
\end{equation}%
where $\lambda $ is the spectral width of the reservoir, and $\gamma _{0}$
represents the coupling strength. We further postulate that the initial
state of the photon system is given by $\cos (\alpha /2)\left\vert
0\right\rangle +e^{i\phi }\sin (\alpha /2)\left\vert 1\right\rangle .$ By
accounting for the gravitational redshift described in Eq. (\ref{18}) and
tracing out the orthogonal mode, the initial state of the photon system
simplifies to \cite{18}
\begin{equation}
\hat{\rho}_{red}^{A}(0)=\left(
\begin{array}{cc}
\sin ^{2}\left( \frac{\alpha }{2}\right) \cos ^{2}\theta & \frac{1}{2}%
e^{i\phi }\sin \alpha \cos \theta \\
\frac{1}{2}e^{-i\phi }\sin \alpha \cos \theta & \cos ^{2}\left( \frac{\alpha
}{2}\right) +\sin ^{2}\left( \frac{\alpha }{2}\right) \sin ^{2}\theta%
\end{array}%
\right) .  \label{24}
\end{equation}%
The corresponding time-dependent reduced density matrix of the photon system
is described as \cite{18}
\begin{equation}
\hat{\rho}_{red}^{A}(t)=\left(
\begin{array}{cc}
P_{t}\sin ^{2}\left( \frac{\alpha }{2}\right) \cos ^{2}\theta & \frac{1}{2}%
\sqrt{P_{t}}e^{i\phi }\sin \alpha \cos \theta \\
\frac{1}{2}\sqrt{P_{t}}e^{-i\phi }\sin \alpha \cos \theta & 1-P_{t}\sin
^{2}\left( \frac{\alpha }{2}\right) \cos ^{2}\theta%
\end{array}%
\right) ,  \label{25}
\end{equation}%
with $P_{t}=e^{-\int_{0}^{t}dt^{\prime }\gamma _{t^{\prime }}}.$ Since the
reservoir is characterized by an effective Lorentzian spectral density
outlined in Eq. (\ref{18}), the time-dependent decay rate adopts the
following form
\begin{equation}
\gamma _{t}=\frac{2\lambda \gamma _{0}\sinh (dt/2)}{d\cosh (dt/2)+\lambda
\sinh (dt/2)},  \label{26}
\end{equation}%
where $d=\sqrt{\lambda ^{2}-2\gamma _{0}\lambda }.$ Based on the Eq. (\ref%
{26}), one can further obtain%
\begin{equation}
P_{t}=e^{-\lambda t}[\cosh (dt/2)+\lambda /d\sinh (dt/2)].  \label{27}
\end{equation}%
It is worth noting that under weak-coupling conditions ($\lambda >2\gamma
_{0}$), the photon system generally follows Markovian dynamics, displaying
irreversible decay. In contrast, under strong-coupling conditions ($\lambda
<2\gamma _{0}$), the system exhibits non-Markovian behavior, marked by the
backflow of information from the environment.

Then, we first consider a two-parameter estimation involving the initial
weight parameter $\alpha $ and phase parameter $\phi $ for the
time-dependent reduced density matrix of the photon system. The
corresponding commutation relation between SLD operators $L_{\alpha }^{S}$
and $L_{\phi }^{S}$ and the mean Uhlmann curvature matrix $D$ can be
respectively expressed as
\begin{eqnarray}
\lbrack L_{\alpha }^{S},L_{\phi }^{S}] &=&2i\left(
\begin{array}{cc}
\Delta _{1} & e^{i\phi }\Delta _{2} \\
e^{-i\phi }\Delta _{2} & -\Delta _{1}%
\end{array}%
\right) ,  \notag \\
D &=&\left(
\begin{array}{cc}
0 & \Delta _{1}\Delta _{3} \\
-\Delta _{1}\Delta _{3} & 0%
\end{array}%
\right) ,  \label{28}
\end{eqnarray}%
where
\begin{eqnarray}
\Delta _{1} &=&P_{t}\sin \alpha \cos ^{2}\theta ,  \notag \\
\Delta _{2} &=&2\sqrt{P_{t}}\cos ^{2}(\alpha /2)\cos \theta ,  \notag \\
\Delta _{3} &=&\cos \alpha +2P_{t}\sin ^{2}(\alpha /2)\cos ^{2}\theta .
\label{29}
\end{eqnarray}%
These results unequivocally reveal that the SLD-CRB is not achievable, not
even in the limit of measurements on an asymptotically large number of
copies of the photon system. Further, by exploiting Eqs. (\ref{3}), (\ref{4}%
), and (\ref{5}), we can derive the SLD-CRB $C_{(\alpha ,\phi )}^{S}$ and
the RLD-CRB $C_{(\alpha ,\phi )}^{R}$%
\begin{eqnarray}
C_{(\alpha ,\phi )}^{S} &=&\frac{1}{P_{t}}(1+\csc ^{2}\alpha )\sec
^{2}\theta ,  \notag \\
C_{(\alpha ,\phi )}^{R} &=&1+\frac{1}{P_{t}}\csc ^{2}\alpha \sec ^{2}\theta
\notag \\
&&+\frac{1}{P_{t}}\left\vert \Lambda \right\vert \sec ^{2}\theta \tan
(\alpha /2),  \label{30}
\end{eqnarray}%
with $\Lambda $=$\csc ^{2}(\alpha /2)+2P_{t}\cos ^{2}\theta -2.$ In general,
the calculation of the HCRB and the NB involves solving a minimization
problem formulated as a SDP, which is computationally non-trivial.
Nevertheless, for two-parameter estimation in a single-qubit system, we can
obtain the corresponding HCRB $C_{(\alpha ,\phi )}^{H}$ and NB $C_{(\alpha
,\phi )}^{N}$ based on analytic expressions derived in Refs. \cite{45,67}%
\begin{eqnarray}
C_{(\alpha ,\phi )}^{H} &=&\left\{
\begin{array}{c}
C_{(\alpha ,\phi )}^{R},\text{ }C_{(\alpha ,\phi )}^{R}\geq \frac{C_{(\alpha
,\phi )}^{S}+C_{(\alpha ,\phi )}^{Z}}{2} \\
C_{(\alpha ,\phi )}^{R}+S_{(\alpha ,\phi )},\text{ }C_{(\alpha ,\phi )}^{R}<%
\frac{C_{(\alpha ,\phi )}^{S}+C_{(\alpha ,\phi )}^{Z}}{2}%
\end{array}%
\right. ,  \notag \\
C_{(\alpha ,\phi )}^{N} &=&C_{(\alpha ,\phi )}^{S}+\frac{2}{P_{t}}\left\vert
\csc \alpha \right\vert \sec ^{2}\theta ,  \label{31}
\end{eqnarray}%
where%
\begin{eqnarray}
C_{(\alpha ,\phi )}^{Z} &=&C_{(\alpha ,\phi )}^{S}+2[\tan (\alpha /2)+\left.
\cot \alpha \sec ^{2}\theta \right/ P_{t}],  \notag \\
S_{(\alpha ,\phi )} &=&\frac{\left[ \left. (C_{(\alpha ,\phi
)}^{Z}+C_{(\alpha ,\phi )}^{S})\right/ 2-C_{(\alpha ,\phi )}^{R}\right] ^{2}%
}{C_{(\alpha ,\phi )}^{Z}-C_{(\alpha ,\phi )}^{R}}.  \label{32}
\end{eqnarray}

\begin{figure}[tbp]
\label{Fig2} \centering\includegraphics[width=0.9\columnwidth]{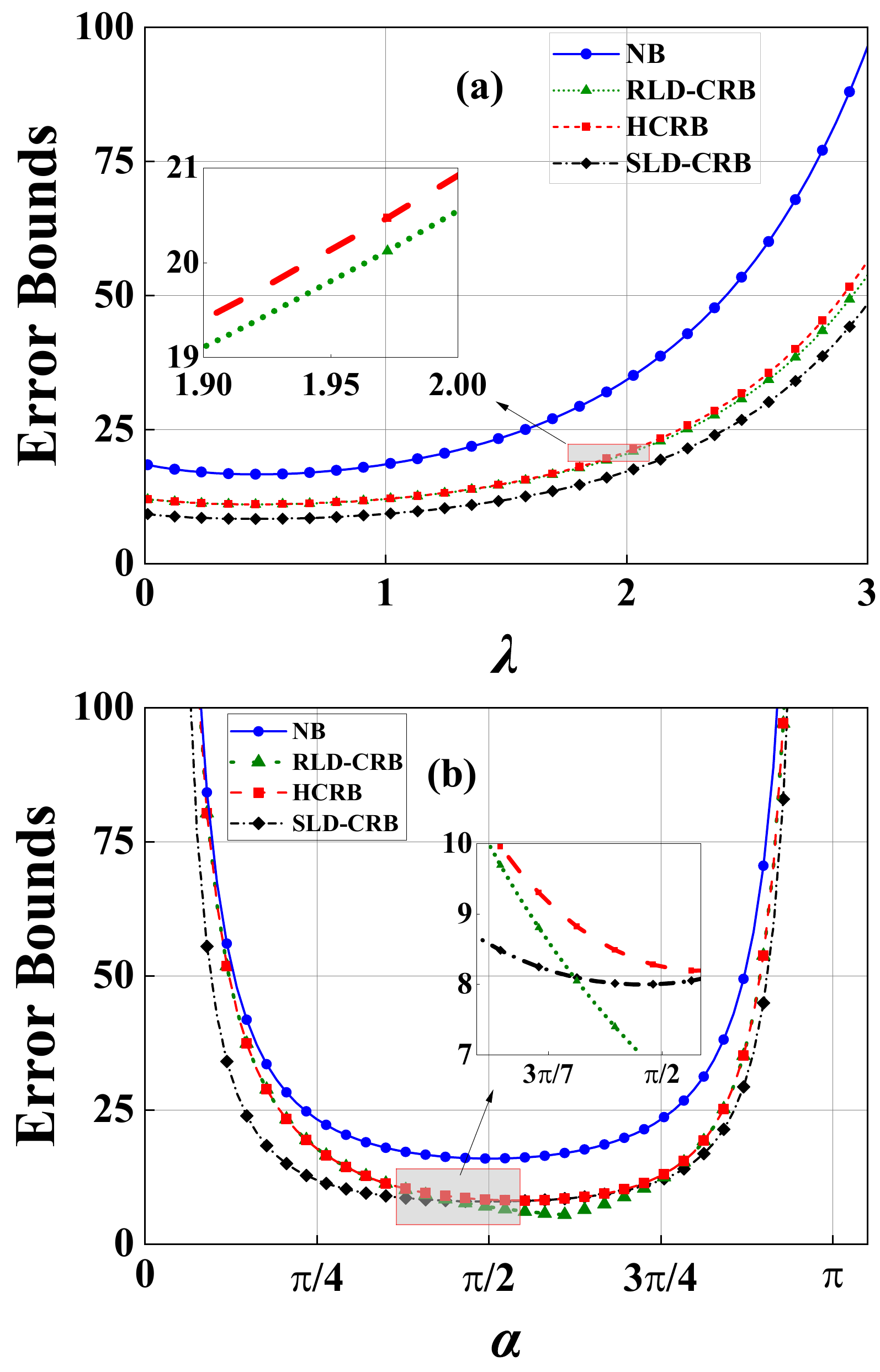}%
\newline
\caption{{}(Color online) Error bounds as a function of (a) the spectral
width $\protect\lambda $ of the reservoir\ with $\protect\alpha =\protect%
\theta =\protect\pi /3$ and\ $\protect\gamma _{0}=t=1,$ and of (b) the
initial weight parameter $\protect\alpha $\ with $\protect\theta =\protect%
\pi /3$ and\ $\protect\lambda =\protect\gamma _{0}=t=1.$}
\end{figure}

To comprehensively evaluate the estimation performance of the SLD-CRB,
RLD-CRB, HCRB, and NB, we systematically analyze their dependence on the
spectral width $\lambda $ of the reservoir, as illustrated in Fig. 2(a). The
results demonstrate a monotonic increase in these error bounds with
expanding $\lambda $ values. Notably, the RLD-CRB and HCRB are nearly
identical in numerical value and consistently exceed the SLD-CRB, suggesting
that both RLD-CRB and HCRB serve as asymptotically tight precision limits,
while the SLD-CRB exhibits the poorest tightness. Furthermore, the NB
consistently yields the largest values, substantiating its role as the most
fundamental precision limit in this multiparameter setting. Comparative
analysis reveals that for $\gamma _{0}=1,$\textbf{\ }the attainable error
bounds (HCRB and NB) exhibit superior estimation precision in the
strong-coupling regime ($\lambda <2$) compared to the weak-coupling regime ($%
\lambda >2$). Further, we also consider the effects of the initial weight
parameter $\alpha $\ on the estimation performance of these error bounds, as
shown in Fig. 2(b). It is shown that these error bounds exhibit a
non-monotonic trend, initially decreasing and then increasing with
variations in the initial weight parameter $\alpha .$\ Specifically, for $%
\alpha <0.464,$\ the RLD-CRB, HCRB and NB are nearly identical, indicating
that these error bounds can provide tight precision limits. In the interval $%
0.464<\alpha <1.371,$\ the RLD-CRB and HCRB are nearly the same and exceed
the SLD-CRB, suggesting that both RLD-CRB and HCRB serve as asymptotically
tight precision limits, while the SLD-CRB exhibits the poorest tightness. As
$\alpha $\ increases further into the range of $1.371<\alpha <2.325,$\ the
SLD-CRB and HCRB converge and exceed the RLD-CRB. Finally, for $\alpha >2.325
$, the RLD-CRB and HCRB again align, outperforming the SLD-CRB.

\begin{figure}[tbp]
\label{Fig3} \centering\includegraphics[width=0.9\columnwidth]{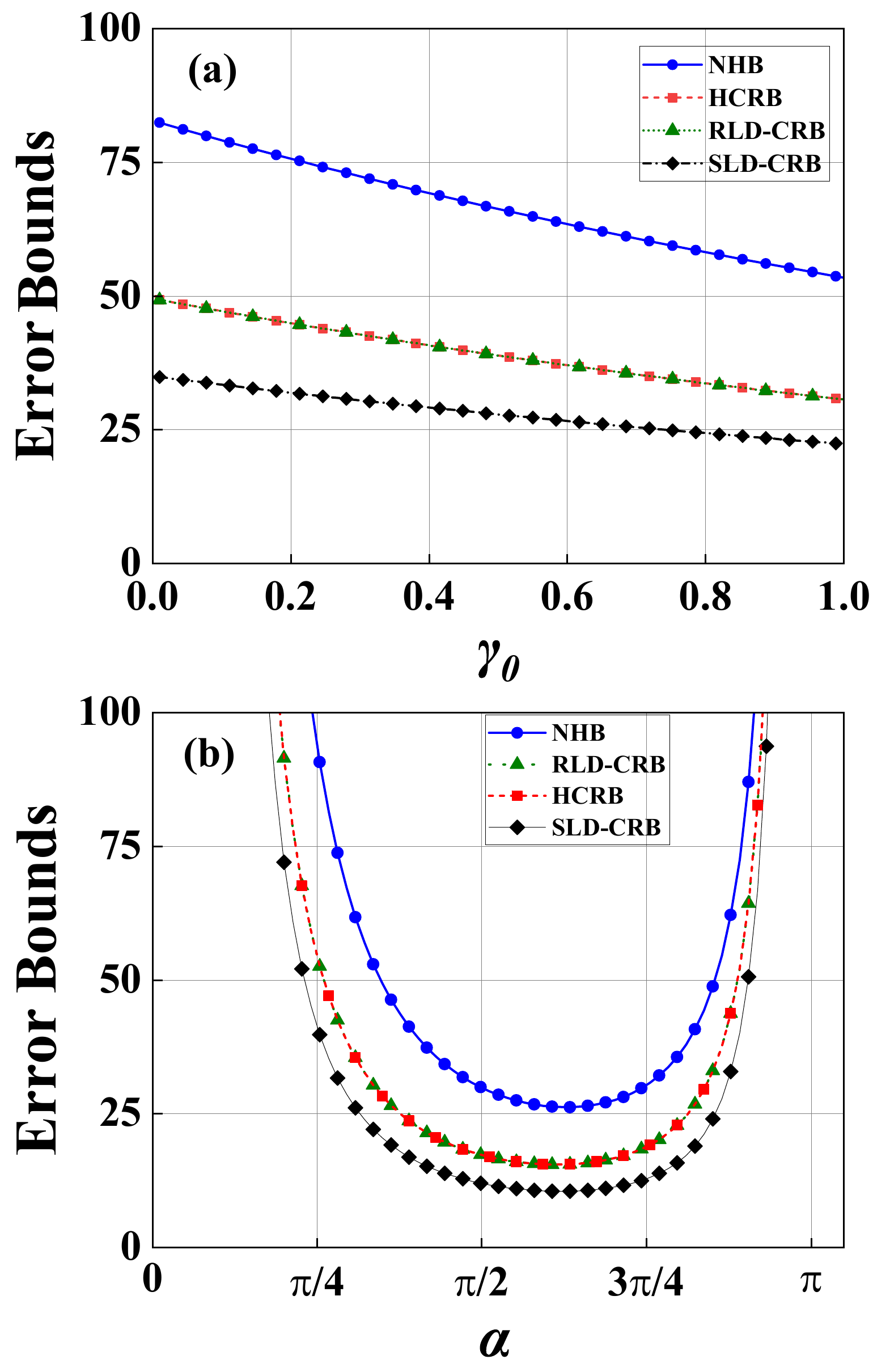}%
\newline
\caption{{}(Color online) Error bounds as a function of (a) the $\protect%
\gamma _{0}$\ with $\protect\alpha =\protect\theta =\protect\pi /3$, and $%
\protect\lambda =t=1,$ and of (b) the $\protect\alpha $\ with $\protect%
\theta =\protect\pi /3$ and\ $\protect\lambda =\protect\gamma _{0}=t=1.$}
\end{figure}

Next, we examine a three-parameter estimation problem, including the initial
weight parameter $\alpha $, the phase parameter $\phi $, and the strength $%
\theta $ of gravitational redshift. Due to the complexity of the analytical
results, we focus on numerical comparisons of SLD-CRB, RLD-CRB, HCRB, and
NHB as functions of the relevant physical parameters, as shown in Fig. 3.
Notably, the NHB consistently yields the largest values among all the
bounds, confirming its role as the tightest achievable precision limit.
Furthermore, the RLD-CRB and HCRB are numerically identical and both exceed
the SLD-CRB, indicating that while the RLD-CRB and HCRB provide
asymptotically tight precision limits, the SLD-CRB exhibits the weakest
tightness. Our analysis reveals that for $\lambda =1,$\ the attainable error
bounds (HCRB and NHB) demonstrate superior estimation precision in the
strong-coupling regime ($\gamma _{0}>0.5$) compared to the weak-coupling
regime ($\gamma _{0}<0.5$).

\subsection{Ohmic-like dephasing channel}

Next, we examine the interaction between a photon system and a bosonic
environment with the Hamiltonian \cite{18,54}
\begin{equation}
\hat{H}=\omega _{0}\hat{a}^{\dagger }\hat{a}+\sum\limits_{k}[\omega _{k}\hat{%
b}_{k}^{\dagger }\hat{b}_{k}+\hat{a}^{\dagger }\hat{a}(g_{k}\hat{b}%
_{k}+g_{k}^{\ast }\hat{b}_{k}^{\dagger })],  \label{33}
\end{equation}%
where $\omega _{0}$ and $\omega _{k}$ are the transition frequency of the
photon system and the $k$th field frequency of the environmental mode,
respectively. We assume that there exists no initial correlation between the
system and the environment, and that the environment starts in a
zero-temperature vacuum state. The dynamics of the system can be accurately
described using the Ohmic-like dephasing model. The nonunitary generator of
the reduced dynamics of the system can be expressed as \cite{18,54}
\begin{equation}
\text{\L }_{O}(\hat{\rho}_{t})=\frac{\gamma _{t}}{2}[2\hat{a}^{\dagger }\hat{%
a}\hat{\rho}_{t}\hat{a}^{\dagger }\hat{a}-(\hat{a}^{\dagger }\hat{a})^{2}%
\hat{\rho}_{t}-\hat{\rho}_{t}(\hat{a}^{\dagger }\hat{a})^{2}].  \label{34}
\end{equation}%
The corresponding bosonic environment operator is characterized as a linear
coupling of the harmonic-oscillator continuum coordinates, which is governed
by the spectral function $J(\omega ).$ The time-dependen decay rate $\gamma
_{t}$ takes the form%
\begin{equation}
\gamma _{t}=\int_{0}^{\infty }J(\omega )\coth \left( \frac{\text{%
h{\hskip-.2em}\llap{\protect\rule[1.1ex]{.325em}{.1ex}}{\hskip.2em}%
}\omega }{2K_{B}T}\right) \frac{1-\cos (\omega t)}{\omega ^{2}}d\omega ,
\label{35}
\end{equation}%
where $K_{B}$ and $T$ denotes the Boltzmann constant and the temperature,
respectively. Further posit that the spectral density of the environment
modes satisfy the Ohmic-like%
\begin{equation}
J(\omega )=\eta \frac{\omega ^{s}}{\omega _{c}^{s-1}}e^{-\frac{\omega }{%
\omega _{c}}},  \label{36}
\end{equation}%
where $\eta $ is a dimensionless coupling constant and $\omega _{c}$
represents the cutoff frequency. By varying the numerical values of the $s$%
-parameter, one can obtain three types of environments, i.e., sub-Ohmic ($%
0<s<1),$ Ohmic ($s=1$), and super-Ohmic ($s>1$). In general, when the
reservoir spectrum is super-Ohmic, memory effects emerge, giving rise to
phenomena such as information backflow and recoherence \cite{68,69}. In
particular, in the case of environment temperature $T=0,$ $t>0$ and $s>0,$
the decay rate $\gamma _{t}$ is given by%
\begin{equation}
\gamma _{t}=\eta \left[ 1-\frac{\cos [(s-1)\arctan (\omega _{c}t)]}{%
(1+\omega _{c}^{2}t^{2})^{\frac{s-1}{2}}}\right] \Gamma (s-1),  \label{37}
\end{equation}%
where $\Gamma (s-1)$ is the Euler gamma function. When the parameter $s$
approaches $1,$ the decay rate $\gamma _{t}$ simplifies to $\gamma _{t}=\eta
\ln (1+\omega _{c}^{2}t^{2}).$ Likewise, we choose the initial state of the
photon system as $\cos (\alpha /2)\left\vert 0\right\rangle +e^{i\phi }\sin
(\alpha /2)\left\vert 1\right\rangle .$ By incorporating the gravitational
redshift effect as delineated in Eq. (\ref{18}) and tracing out the
orthogonal mode, the corresponding time-dependent reduced density matrix of
the photon system is interpreted as \cite{18}
\begin{equation}
\hat{\rho}_{red}^{O}(t)=\left(
\begin{array}{cc}
\sin ^{2}\left( \frac{\alpha }{2}\right) \cos ^{2}\theta & \frac{1}{2}%
q_{t}e^{i\phi }\sin \alpha \cos \theta \\
\frac{1}{2}q_{t}e^{-i\phi }\sin \alpha \cos \theta & \cos ^{2}\left( \frac{%
\alpha }{2}\right) +\sin ^{2}\left( \frac{\alpha }{2}\right) \sin ^{2}\theta%
\end{array}%
\right) ,  \label{38}
\end{equation}%
where $q_{t}=e^{-\gamma _{t}/2}.$

Based on Eq. (\ref{38}), we first consider a two-parameter estimation
including the initial weight parameter $\alpha $ and the phase parameter $%
\phi .$ The associated commutation relations between the SLD operators $%
L_{\alpha }^{S}$ and $L_{\phi }^{S},$ as well as the mean Uhlmann curvature
matrix $D,$ can be articulated as follow%
\begin{eqnarray}
\lbrack L_{\alpha }^{S},L_{\phi }^{S}] &=&i\left(
\begin{array}{cc}
\left. -\Theta _{2}\right/ \Theta _{1} & \left. e^{i\phi }\Theta _{3}\Theta
_{4}\right/ \Theta _{1} \\
\left. e^{-i\phi }\Theta _{3}\Theta _{4}\right/ \Theta _{1} & \left. \Theta
_{2}\right/ \Theta _{1}%
\end{array}%
\right)  \notag \\
D &=&\left(
\begin{array}{cc}
0 & \Theta _{5} \\
-\Theta _{5} & 0%
\end{array}%
\right)  \label{39}
\end{eqnarray}%
where%
\begin{eqnarray}
\Theta _{1} &=&\cos (2\theta )+2q_{t}-3  \notag \\
&&-[1-2q_{t}^{2}+\cos (2\theta )]\cos \alpha ,  \notag \\
\Theta _{2} &=&2q_{t}^{2}\sin ^{2}(2\theta )\sin ^{2}(\alpha /2)\sin \alpha ,
\notag \\
\Theta _{3} &=&4q_{t}\cos ^{2}(\alpha /2)\cos \theta ,  \notag \\
\Theta _{4} &=&q_{t}^{2}[1+\cos \alpha +2\cos (2\theta )\sin ^{2}(\alpha
/2)]-2,  \notag \\
\Theta _{5} &=&q_{t}^{2}[\cos ^{4}\theta \sin \alpha +\left. \sin
^{2}(2\theta )\sin (2\alpha )\right/ 8].  \label{40}
\end{eqnarray}%
From the Eq. (\ref{39}), we can clearly see that the SLD-CRB is not a tight
error bound. Then, by using Eqs. (\ref{3}), (\ref{4}), and (\ref{5}), we can
obtain the corresponding SLD-CRB $C_{(\alpha ,\phi )}^{S}$ and the RLD-CRB $%
C_{(\alpha ,\phi )}^{R}$%
\begin{eqnarray}
C_{(\alpha ,\phi )}^{S} &=&(\left. \Xi _{1}\right/ \Xi _{2}+\left. \csc
^{2}\alpha \right/ q_{t}^{2})\sec ^{2}\theta ,  \notag \\
C_{(\alpha ,\phi )}^{R} &=&\left. \Xi _{1}\Xi _{4}\right/ 2q_{t}^{2}\Xi
_{2}-\left. \Xi _{3}\right/ \Xi _{2}+\left\vert \left. \Xi _{5}\Xi
_{6}\right/ \Xi _{7}\right\vert ,  \label{41}
\end{eqnarray}%
where%
\begin{eqnarray}
\Xi _{1} &=&3-2q_{t}^{2}+\cos \alpha -2q_{t}^{2}\cos \alpha  \notag \\
&&-(1-\cos \alpha )\cos (2\theta ),  \notag \\
\Xi _{2} &=&2-q_{t}^{2}+(2-3q_{t}^{2})\cos \alpha  \notag \\
&&-q_{t}^{2}(1-\cos \alpha )\cos (2\theta ),  \notag \\
\Xi _{3} &=&2-2\cos \alpha +2(q_{t}^{2}+q_{t}^{2}\cos \alpha -2)\sec
^{2}\theta ,  \notag \\
\Xi _{4} &=&[1-2q_{t}^{2}+\cos (2\theta )+2q_{t}^{2}\csc ^{2}\alpha ]\sec
^{2}\theta ,  \notag \\
\Xi _{5} &=&[1-2q_{t}^{2}+\cos (2\theta )-2\left( 1-q_{t}^{2}\right) \csc
^{2}(\alpha /2)]\sec ^{2}\theta ,  \notag \\
\Xi _{6} &=&2[\cot ^{2}(\alpha /2)+\cos (2\theta )]\tan (\alpha /2),  \notag
\\
\Xi _{7} &=&4-6q_{t}^{2}+2q_{t}^{2}\cos (2\theta )-4(1-q_{t}^{2})\csc
^{2}(\alpha /2).  \label{42}
\end{eqnarray}%
Likewise, we further get the corresponding HCRB $C_{(\alpha ,\phi )}^{H}$
and NB $C_{(\alpha ,\phi )}^{N}$ based on Refs. \cite{45,67}%
\begin{eqnarray}
C_{(\alpha ,\phi )}^{H} &=&\left\{
\begin{array}{c}
C_{(\alpha ,\phi )}^{R},\text{ }C_{(\alpha ,\phi )}^{R}\geq \frac{C_{(\alpha
,\phi )}^{S}+C_{(\alpha ,\phi )}^{Z}}{2} \\
C_{(\alpha ,\phi )}^{R}+S_{(\alpha ,\phi )},\text{ }C_{(\alpha ,\phi )}^{R}<%
\frac{C_{(\alpha ,\phi )}^{S}+C_{(\alpha ,\phi )}^{Z}}{2}%
\end{array}%
\right. ,  \notag \\
C_{(\alpha ,\phi )}^{N} &=&C_{(\alpha ,\phi )}^{S}+2\sqrt{\left. \Lambda
\right/ q_{t}^{2}\Xi _{2}},  \label{43}
\end{eqnarray}%
where%
\begin{eqnarray}
C_{(\alpha ,\phi )}^{Z} &=&C_{(\alpha ,\phi )}^{S}+2\left\vert \left. \Xi
_{1}(1+\cos \alpha \tan ^{2}\theta )\csc \alpha \right/ \Xi _{2}\right\vert ,
\notag \\
S_{(\alpha ,\phi )} &=&\frac{\left[ \left. (C_{(\alpha ,\phi
)}^{Z}+C_{(\alpha ,\phi )}^{S})\right/ 2-C_{(\alpha ,\phi )}^{R}\right] ^{2}%
}{C_{(\alpha ,\phi )}^{Z}-C_{(\alpha ,\phi )}^{R}},  \notag \\
\Lambda &=&\csc \alpha \sec ^{4}\theta \lbrack \cot (\alpha
)(1-2q_{t}^{2}+\cos (2\theta ))  \notag \\
&&+\csc (\alpha )(3-2q_{t}^{2}-\cos (2\theta ))].  \label{44}
\end{eqnarray}

\begin{figure}[tbp]
\label{Fig4} \centering\includegraphics[width=0.9\columnwidth]{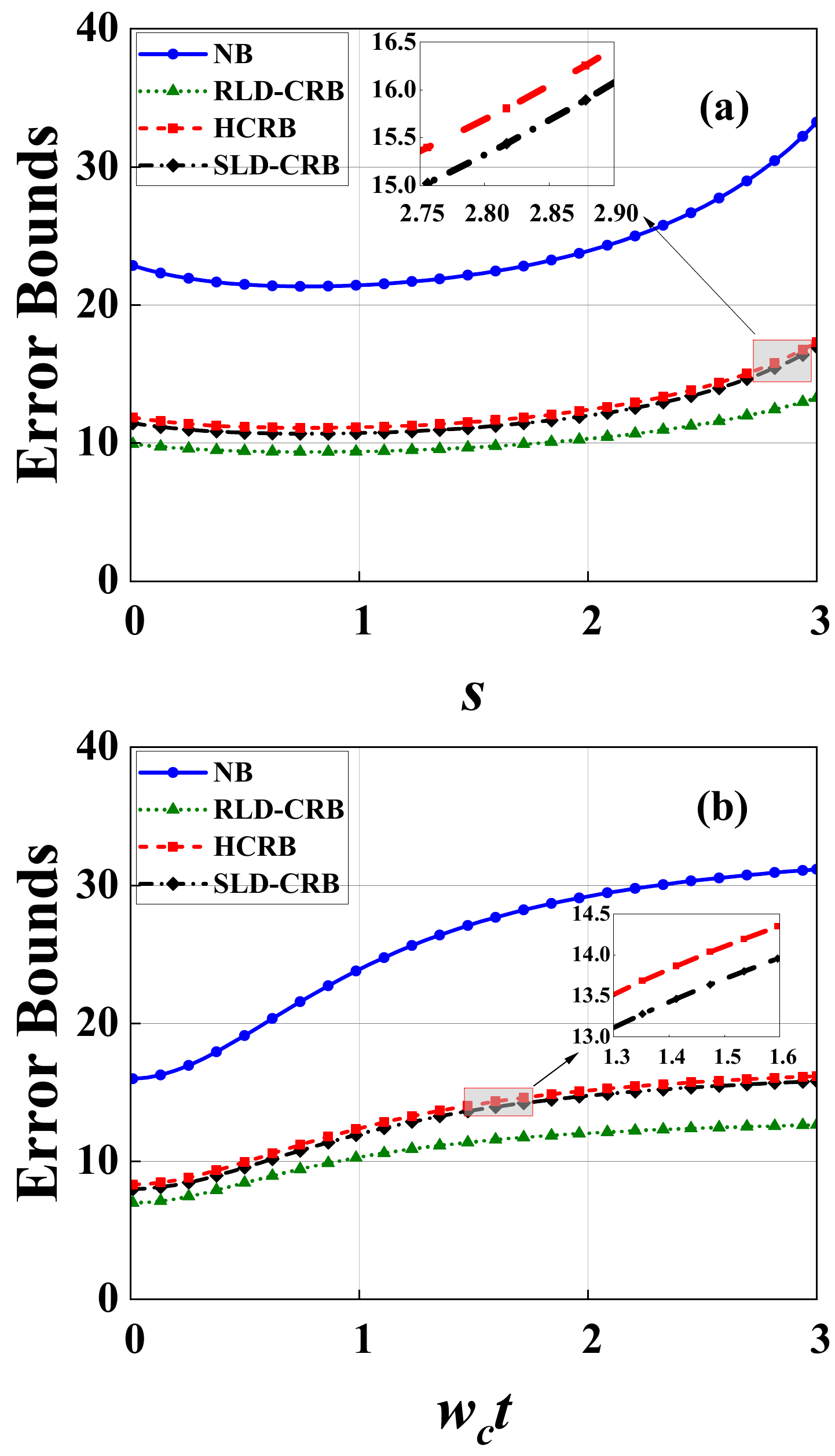}%
\newline
\newline
\caption{{}(Color online) Error bounds as a function of (a) the $s$%
-parameter\ with $\protect\alpha =\protect\pi /2$, $\protect\theta =\protect%
\pi /3,$ and $\protect\omega _{c}=t=\protect\eta =1,$ and of (b) the time
parameter $\protect\omega _{c}t$\ with $\protect\alpha =\protect\pi /2$, $%
\protect\theta =\protect\pi /3,$ $s=2,$ and $\protect\eta =1.$}
\end{figure}
To assess the estimation performance of these error bounds, we plotted them
as a function of the $s$-parameter, as shown in Fig. 4(a). The results
indicate that as the value of $s$ increases, these error bounds gradually
increase numerically. Notably, the SLD-CRB and HCRB are nearly identical and
both exceed the RLD-CRB, indicating that while the SLD-CRB and HCRB provide
asymptotically tight precision limits, the RLD-CRB exhibits the weakest
tightness. Furthermore, a comparative analysis of different spectral regimes
reveals that the attainable error bounds (HCRB and NB) manifest
significantly enhanced estimation precision in the sub-Ohmic regime ($0<s<1$%
) compared to both the Ohmic ($s=1$) and super-Ohmic ($s>1$) regimes.
Crucially, the NB bound consistently maintains the highest numerical values
across all examined cases, thereby unequivocally establishing the tightest
precision limit among the considered bounds. Additionally, we investigate
the influence of the time parameter $\omega _{c}t$\ on the estimation
performance of these error bounds, as depicted in Fig. 4(b). The results
demonstrate that these error bounds exhibit a gradual numerical increase
with respect to $\omega _{c}t$, ultimately converging to a stable asymptotic
value.
\begin{figure}[tbp]
\label{Fig5} \centering\includegraphics[width=0.9\columnwidth]{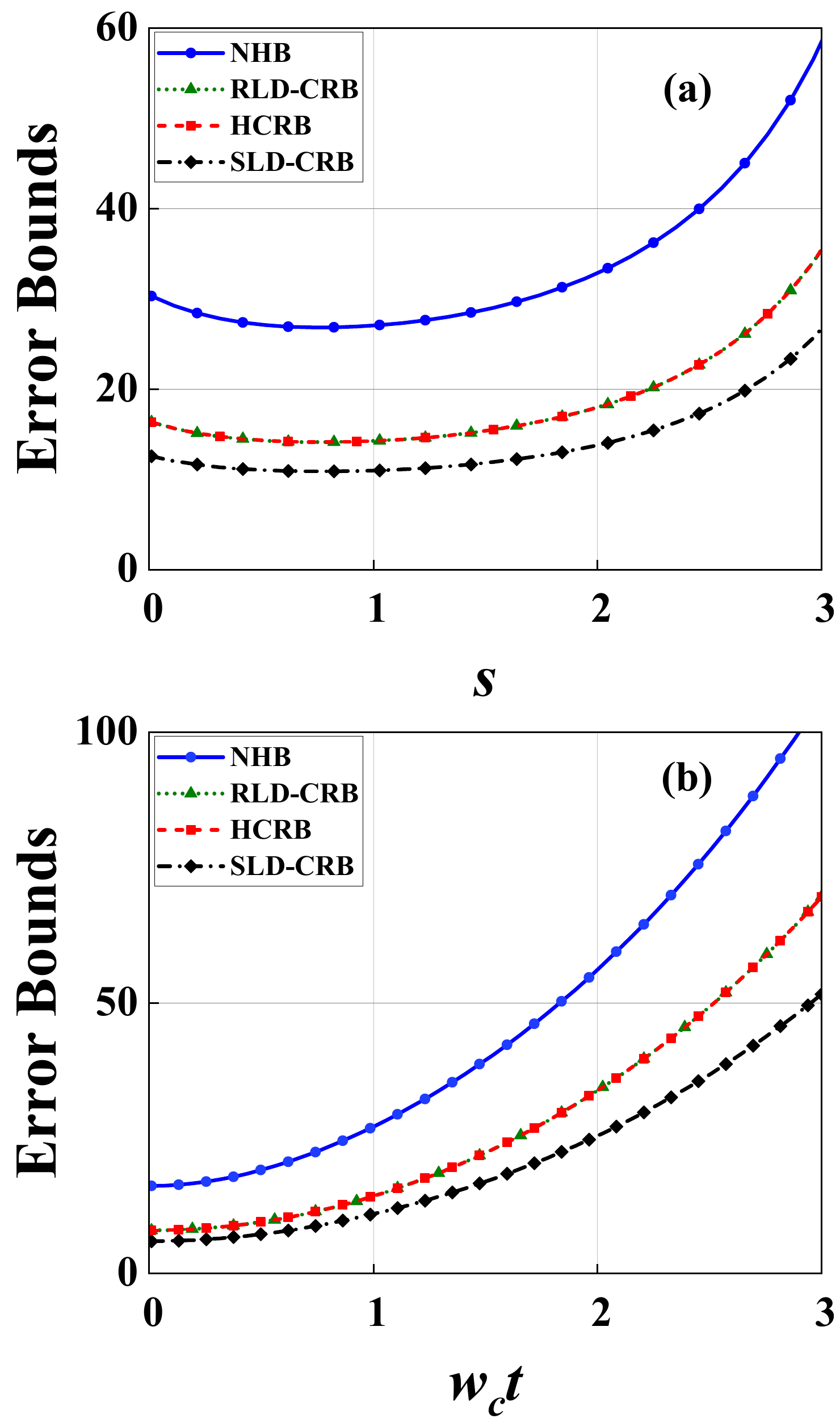}%
\newline
\newline
\caption{{}(Color online) Error bounds as a function of (a) the $s$%
-parameter\ with $\protect\alpha =\protect\pi /2$, $\protect\theta =\protect%
\pi /4,$ and $\protect\omega _{c}=t=\protect\eta =1,$ and of (b) the time
parameter $\protect\omega _{c}t$\ with $\protect\alpha =\protect\pi /2$, $%
\protect\theta =\protect\pi /4,$ $s=0.5,$ and $\protect\eta =1.$}
\end{figure}

Subsequently, we investigate a three-parameter estimation problem, including
the initial weight parameter $\alpha $, the phase parameter $\phi $, and the
gravitational redshift strength $\theta $. Given the complexity of the
analytical results, our focus shifts to a numerical comparison of the
SLD-CRB, RLD-CRB, HCRB, and NHB as functions of the pertinent physical
parameters, as illustrated in Fig. 5. Our findings reveal that the
estimation performance of the attainable error bounds (HCRB and NHB) is
enhanced in the sub-Ohmic regime ($0<s<1$) when compared to both the Ohmic ($%
s=1$) and super-Ohmic ($s>1$) scenarios, as shown in Fig. 5(a). Notably, the
NHB consistently yields the largest values among all the bounds, confirming
its role as the tightest achievable precision limit. The RLD-CRB is
numerically equivalent to the HCRB, indicating that both provide
asymptotically tight precision limits. Furthermore, these error bounds
demonstrate a numerical increase with respect to $\omega _{c}t$\ in Fig.
5(b).

\section{Conclusions}

In conclusion, we have investigated multiparameter quantum estimation for a
photon system subjected to amplitude-damping and Ohmic-like dephasing
channels under the effect of gravitational redshift. For the
amplitude-damping channel, we first examine a two-parameter estimation
problem involving the initial weight and phase parameters, deriving
analytical expressions for the SLD-CRB, RLD-CRB, HCRB, and NB. Our findings
reveal that the attainable error bounds (HCRB and NB) achieve significantly
higher estimation precision in the strong-coupling regime compared to the
weak-coupling regime. Notably, the NB yields the tightest error bound among
all bounds, consistent with the general hierarchy of multiparameter quantum
estimation. Subsequently, we explore a three-parameter estimation problem
incorporating the weight parameter, phase parameter, and gravitational
redshift strength. Given the complexity of the analytical results, we focus
on numerical comparisons of the SLD-CRB, RLD-CRB, HCRB, and NHB using SDP.
These comparisons yield similar observations: the estimation performance of
the attainable error bounds (HCRB and NHB) is consistently superior under
strong-coupling conditions relative to weak-coupling scenarios. For the
Ohmic-like dephasing channel, we also analyze both two-parameter and
three-parameter estimation. Our results demonstrate that the attainable
error bounds (HCRB, NB, and NHB) consistently exhibit enhanced estimation
precision in the sub-Ohmic regime compared to the Ohmic and super-Ohmic
regimes. Intriguingly, for both the amplitude-damping and Ohmic-like
dephasing channels, the RLD-CRB in the three-parameter estimation is
numerically equivalent to the HCRB, indicating both RLD-CRB and HCRB can
provide asymptotically tight precision limits.

\begin{acknowledgments}
We sincerely thank Marco G. Genoni, Francesco Albarelli, and Simon K. Yung
for helpful discussions. This work is supported by Jiangxi Provincial
Natural Science Foundation (20252BAC240169), and Ying Xia is supported by Shaanxi Association for Science and Technology Young Talent Support Program (20260521).
\end{acknowledgments}

\end{document}